\documentclass[review]{elsarticle}

\usepackage{lineno,hyperref}
\modulolinenumbers[5]
\usepackage[utf8]{inputenc}
\usepackage[english]{babel}
\usepackage{amsmath,amssymb}
\usepackage{textcomp}
\usepackage{mathtools}
\usepackage{longtable}
\usepackage{float}
\usepackage{lscape}
\usepackage{graphicx}
\usepackage{booktabs}
\usepackage[section]{placeins}
\journal{Journal of Molecular Spectroscopy}

\renewcommand{\thefootnote}{\alph{footnote}}

\newcommand{\braket}[1]{\left< #1\right>} % for Dirac brackets

\begin{document}

\begin{frontmatter}

\title{Structure and dynamics of H$_2^+$ near the dissociation threshold: a combined experimental and computational investigation}

\author{Maximilian Beyer}
\author{Fr\'ed\'eric Merkt\corref{mycorrespondingauthor}}
\address{Laboratory of Physical Chemistry, ETH Zurich, CH-8093 Zurich, Switzerland}
\cortext[mycorrespondingauthor]{merkt@phys.chem.ethz.ch}

\begin{abstract}
The pulsed-field-ionization zero-kinetic-energy photoelectron spectrum of H$_2$ has been recorded in the vicinity of the dissociative-ionization threshold following three-photon excitation via selected rotational levels of the B $^1\Sigma_u^+$~($v=19$) and $\bar{\rm H}$ $^1\Sigma_g^+$~($v=11$) intermediate states. The spectra consist of transitions to bound levels of the X$^+$ $^2\Sigma_g^+$ state of H$_2^+$ with $v^+$ in the range 14--19 and $N^+$ in the range 0--9, of the A$^+$ $^2\Sigma_u^+$ state with $v^+=0$ and $N^+=0-2$, and of shape resonances corresponding to the X$^+\ (v^+=17,N^+=7)$ and X$^+\ (v^+=18,N^+=4)$ quasibound levels. Calculations of the level structure of H$_2^+$ have been carried out and the influence of adiabatic, nonadiabatic, relativistic and radiative corrections on the positions of these levels, and in the case of the shape resonances also on their widths, has been investigated. Different methods of calculating the widths and profiles of the shape resonances have been tested for comparison with the experimental observations. Slow oscillations of the dissociative-ionization yield have been observed and reflect, in first approximation, the Franck-Condon factors of the $\bar{\text H}$ $\rightarrow$ X$^+$, A$^+$ bound - free transitions.
\end{abstract}


\end{frontmatter}

\section{Introduction}
H$_2^+$ is the simplest molecule and plays in molecular physics the role that the hydrogen atom plays in atomic physics. The level structure of H$_2^+$ can be calculated with exquisite precision and accuracy by ab-initio quantum-chemical methods, either by solving the eigenvalue problem of the two-proton-one-electron system directly using variational methods \cite{bishop74b,taylor99a,karr06a,korobov06a} and artificial-channel-scattering methods \cite{moss93a}, or by first making an "exact" electronic-structure calculation for clamped nuclei in the realm of the Born-Oppenheimer approximation, and then calculating adiabatic (i.e., electronically diagonal), nonadiabatic (i.e., electronically off-diagonal) and relativistic corrections by perturbation-theoretical methods \cite{peek65b, hunter67a, kolos69a, bishop73a, wolniewicz78a, wolniewicz86a, wolniewicz91a}. Radiative corrections are typically evaluated using the second route \cite{bukowski92a,korobov04a,korobov06b}. From such calculations, an extraordinarily detailed knowledge of the spectrum of H$_2^+$ has resulted. The spectral positions of the low-lying rovibrational levels of the X$^+$ $^2\Sigma_g^+$ ground electronic state have been calculated at a precision of about 2 kHz \cite{karr06a,korobov06a,korobov09a,zhong09a}. The positions of all 423 bound levels of the X$^+$ state and the three bound levels of the A$^+$ $^2\Sigma_u^+$ state ($v^+=0, N^+=0,1$ and 2) are tabulated by Moss \cite{moss93a}. A fourth bound level of the A$^+$ state, with a nonrelativistic binding energy of only $E/h=7.139253$~MHz, was reported later \cite{carbonell03a,carbonell04a}. 

Several energy intervals between fine-structure \cite{jefferts68a}, hyperfine-structure \cite{jefferts69a,osterwalder04a}, rotational \cite{arcuni90a,critchley01a,haase15a}, rovibrational \cite{herzberg72a} and rovibronic energy levels \cite{carrington93b} of H$_2^+$ have also been measured. 

Quasibound levels of H$_2^+$ above the dissociation limit, often called shape resonances, have also been calculated, but less precisely. 58 such resonances are known to exist from the work of Moss \cite{moss93a} who, however, did not report any data on 19 of them, because they are located too close to the maxima of the centrifugal potential barriers for the calculations to reach the desired accuracy. The ten lowest of these resonances have been calculated by Davis and Thorson within the Born-Oppenheimer approximation \cite{davis78a}. Moss has reported adiabatic corrections for these resonances \cite{moss96a} but did not evaluate nonadiabatic corrections nor did he report resonance widths. Recently, we have observed two such shape resonances by PFI-ZEKE photoelectron spectroscopy, corresponding to the ($v^+=17,N^+=7$) and ($v^+=18,N^+=4$) quasibound levels \cite{beyer16a}, in spectra in which several of the bound levels of the X$^+$ and A$^+$ states located near the dissociation limit of H$_2^+$ were also observed. We also calculated the positions and widths of these resonances and found overall good agreement, except for the width of the (17,7) resonance, which the calculations predicted to be narrower than found experimentally. The purpose of the present article is to present new experimental and computational results obtained on a broader range of highly excited levels of H$_2^+$.

Shape resonances arising from the nuclear motion in molecules have been discovered before, and correctly interpreted very soon after, the introduction of the quantum theory \cite{bonhoeffer27a,villars30a}, in relation to what Herzberg classified in 1931 as special case III of predissociation \cite{herzberg31a}, i.e., rotational predissociation observed in the spectra of diatomic molecules such as AlH. Since then, rotational predissociation and the corresponding shape resonances have been observed in a multitude of molecular systems. Their quantitative description requires high-quality potential-energy functions and large progress has been made in the development of reliable computational procedures to determine their positions and widths. The shape resonances of H$_2$ represent an early example for which computations could be performed with spectroscopic accuracy \cite{roy71a}. Today, computer codes such as the program Level \cite{roy14a} are used by experimentalists to calculate resonance widths and positions. 

In the case of H$_2^+$ discussed in this article, achieving spectroscopic accuracy in the computation of shape resonances poses several problems. i) Adiabatic, nonadiabatic, relativistic and radiative corrections are all large enough that they need to be considered in the theoretical treatment. ii) To retain the concept of a potential curve and nevertheless include nonadiabatic corrections, which, per definition, mix different electronic states, requires specific measures, such as the introduction of $R$-dependent reduced masses for the vibrational and rotational motions of the nuclei (see, e.g., \cite{bunker77a,schwenke01a,kutzelnigg07a,jaquet08a}). iii) Protons are light and tunneling resonances can be broad so that Lorentzian line shapes may not always be appropriate to describe them. The approach we followed to calculate the shape resonances was largely inspired by the work of Moss \cite{moss93a}, Wolniewicz and coworkers \cite{wolniewicz78a, wolniewicz91a}, and Kutzelnigg and coworkers \cite{jaquet08a} and consists of (i) evaluating all corrections to the Born-Oppenheimer energy while retaining a single-potential description, and (ii) going beyond semi-classical WKB methods to compute resonance parameters. Because the positions and widths of the resonances might be sensitive to even small variations of the corrections terms, we analyze the effects of the different corrections terms separately and use the comparison with the experimental data and earlier calculations to validate the computation procedure. The validation then enables us to predict the positions and widths of the shape resonances of H$_2^+$ not predicted by Moss in his otherwise complete investigation \cite{moss93a, moss96a}.

\section{Experiment}
\label{experiment}
The bound and quasibound rovibrational levels of H$_2^+$ located near the H$^+$~+~H(1s) dissociation limit were studied by pulsed-field-ionization zero-kinetic-energy (PFI-ZEKE) photoelectron spectroscopy. Because these states have a large average internuclear separation, they are not directly accessible from the X $^1\Sigma_g^+(v=0)$ ground state of H$_2$. A resonant three-photon excitation sequence via the B $^1\Sigma_u^+(v=19,N_{B})$ and $\bar{H}(v=11,N)$ intermediate states were used to gradually enlarge the internuclear separation, as explained in Ref. \cite{beyer16a}. The same sequence also enabled us to select para or ortho H$_2$ by carrying out the excitation through rotational levels of the $\bar{H}$ state with even or odd $N$ value, respectively.

The vacuum-ultraviolet (VUV) radiation with wave number $\tilde\nu_{\rm VUV}=2\tilde\nu_{\rm UV}+\tilde\nu_{\rm 2}$ around 109750~cm$^{-1}$ used to access the B $^1\Sigma_u^+(v=19)$ state from the X $^1\Sigma_g^+(v=0)$ ground state was generated by resonance-enhanced sum-frequency mixing in Kr using two pulsed Nd:YAG-pumped dye lasers (repetition rate 25 Hz, pulse duration 5 ns). The wave number $\tilde\nu_1$ of the first dye laser was tripled using two successive $\beta$-barium-borate crystals and the tripled output ($\tilde\nu_{\rm UV}$) was kept fixed at the position of the $(4p)^55p[3/2](J=0)\leftarrow (4p)^6$ two-photon resonance of Kr ($2\tilde\nu_{\rm UV}=94092.96\ {\rm cm}^{-1}$). The wave number of the second dye laser was then adjusted so as to access the desired rovibrational level of the B state. A third  Nd:YAG-pumped dye laser was used to induce the transition from the selected B state to the $\bar{H}(v=11)$ state, from which the region near the H$_2^+$ dissociation limit was reached with a fourth dye laser, delayed by 10 ns with respect to the other lasers using an optical delay line. This measure helped reducing the intensity of the transitions to Rydberg states of H$_2$ belonging to series converging on the low-lying vibrational levels of H$_2^+$ induced by the fourth laser directly from the B state. 

All laser beams intersected a supersonic expansion of H$_2$ at right angles on the axis of a cylindrically symmetric photoexcitation and electron-extraction region consisting of a set of five parallel and equidistant extraction plates designed for the application of homogeneous electric fields \cite{merkt98a}. Stray magnetic fields were eliminated by a double layer of magnetic schielding and stray electric fields were maintained at a level below 10~mV/cm. The pulsed solenoid valve used to form the supersonic beam was operated at a stagnation pressure of 2 bar of pure H$_2$. The beam was collimated with a skimmer of 1 mm orifice diameter which separated the source chamber from the photoexcitation chamber. The background pressure in the photoexcitation region did not exceed 10$^{-6}$~mbar during operation of the pulsed valve.

The PFI-ZEKE photoelectron spectra of the highest bound levels and the shape resonances of H$_2^+$ were recorded by monitoring the yield of electrons produced by the delayed pulsed field ionization of high-lying Rydberg states (principal quantum number $n>>100$) - either of H$_2$ for the bound levels of H$_2^+$ or of H for the shape resonances - as a function of the wave number of the fourth dye laser. This laser had a bandwith of 0.03~cm$^{-1}$ and its wave number was calibrated at an accuracy of 0.02~cm$^{-1}$ with a wavemeter. The pulsed field ionization was achieved with an electric-field pulse sequence consisting of ten pulses of increasing strength [(1) 50 mV/cm, (2) 70 mV/cm, (3) 80 mV/cm, (4) 90 mV/cm, (5) 110 mV/cm, (6) 140 mV/cm, (7) 200 mV/cm, (8) 270 mV/cm, (9) 680 mV/cm, and (10) 1.22 V/cm)] to selectively detect Rydberg states of progressively lower principal quantum numbers. The electron signals produced by each of these pulses were recorded separately so that each laser scan led to the recording of ten PFI-ZEKE spectra.
The spectrum recorded from pulse (1) turned out to be extremely weak, suffered from undesirable contributions from low-energy electrons, and was therefore not used in the analysis. The best resolution (full width at half maximum of 0.2~cm$^{-1}$) was obtained with pulses (2)-(5). The spectrum recorded with pulse (6) had a resolution of 0.25~cm$^{-1}$, and those recorded with pulse (7) and (8) a resolution of 0.35~cm$^{-1}$. The spectra recorded with pulses (9) and (10) had the best signal-to-noise ratio but the lines in these spectra were too broad to be useful for the determination of line widths and line positions. The relative positions of lines in the photoelectron spectra could be determined at an accuracy of 1 GHz from a statistical analysis of the spectra recorded from the different pulses. The absolute positions of the ionic levels with respect to the selected $\bar{H}$ intermediate levels were determined with an accuracy of about 2~GHz after correcting for the shifts of the ionization thresholds induced by the pulsed electric fields, as described in Reference~\cite{hollenstein01a}.

\section{Experimental results}
\label{exp_results}

\begin{figure}
\includegraphics[width=1.0\textheight, angle=90]{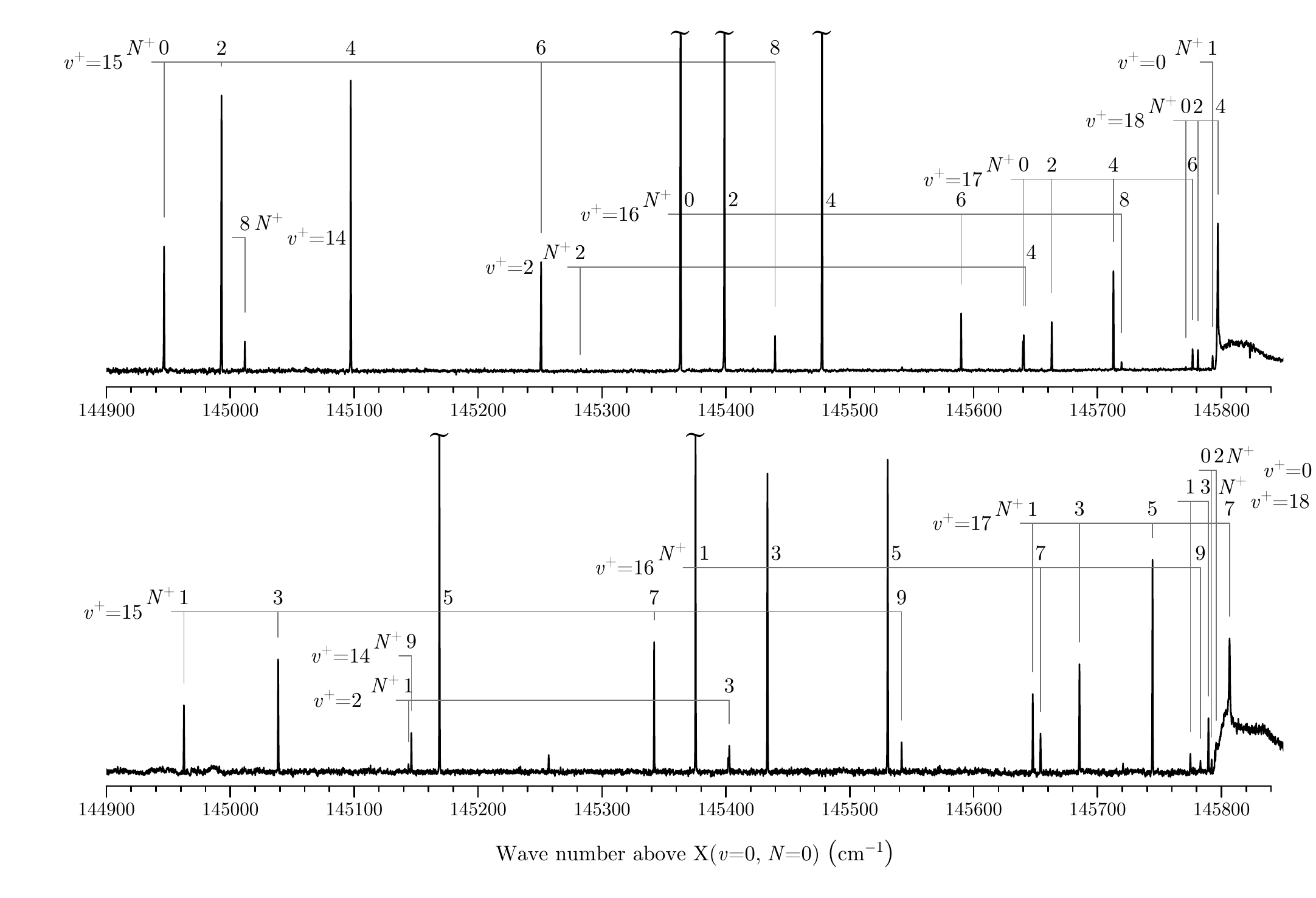}
\caption{PFI-ZEKE photoelectron spectra of H$_2$ recorded near the dissociative-ionization threshold from (a) the $\bar{\rm H}\ (v=11,N=3)$ level and (b) the $\bar{\rm H}\ (v=11,N=2)$ level. The wave numbers are given with respect to the X $(v=0,N=0)$ ground state of H$_2$. The vertical scale is linear and in arbitrary units. The background signal below the dissociative-ionization threshold corresponds to zero signal.}
\label{fig1}
\end{figure}

Overview PFI-ZEKE photoelectron spectra of para and ortho H$_2$ in the region of the dissociation limit of H$_2^+$ (at 145796.84136(37)~cm$^{-1}$ above the X$(v=0,N=0)$ ground state \cite{liu09a,sprecher11a}) are displayed in the upper and lower panels of Fig.~\ref{fig1}, respectively. The spectrum of para H$_2$ was recorded from the $\bar{\rm H}(11,2)$ level and consists of transitions to the A$^+(0,1)$ state and to bound rotational levels of the X$^+(v^+=14-19)$ states with even values of $N^+$ up to 8. The spectrum of ortho H$_2$, recorded from the $\bar{\rm H}(11,3)$ level, reveals transitions to the same vibrational levels of the X$^+$ state but with odd $N^+$ values up to 9 and to the A$^+(0,0)$ and (0,2) states.

The spectra of para and ortho H$_2$ also contain five weak lines that result from transitions to the X$^+(v^+=2, N^+=1-5)$ levels from the B(19,1) and B(19,2) intermediate states. Because the energy difference between the B($v=19,N=1,2$) and the $\bar{\rm H}\ (v=11,N=2,3)$ are precisely known \cite{hinnen94a,reinhold97a,reinhold99a}, the spectrum enables the determination of the relative positions of low- and high-$v^+$ levels of the X$^+$ state (see Table \ref{tab:1} below).

The onset of the dissociation continuum is clearly visible in both spectra and so are the (18,4) and (17,7) shape resonances, which appear as sharp structures in the respective continua. All level positions derived from these spectra, and from similar spectra (not shown) recorded from other rotational levels of the $\bar{\rm H}(v=11)$ state, are given relative to the position of the X$^+$(17,6) level for para H$_2^+$ and the X$^+$(18,3) level for ortho H$_2^+$ in the third column of Table 1. The fourth column of the same table lists the corresponding values of the dissociation energies (positive values for bound levels) derived from the data given in the third column using the dissociation energies of the X$^+$(17,6) (18.5707~cm$^{-1}$ \cite{moss93a}) and X$^+$(18,3) (6.0329~cm$^{-1}$ \cite{moss93a}) levels. For comparison, the dissociation energies calculated by Moss \cite{moss93a} are given in the last column. The table also contains the widths, already reported in Ref.~\cite{beyer16a}, of the (17,7) and (18,4) shape resonances derived from the experimental spectra by deconvolution.
\begin{figure}
\centering
\includegraphics[width=0.7\columnwidth]{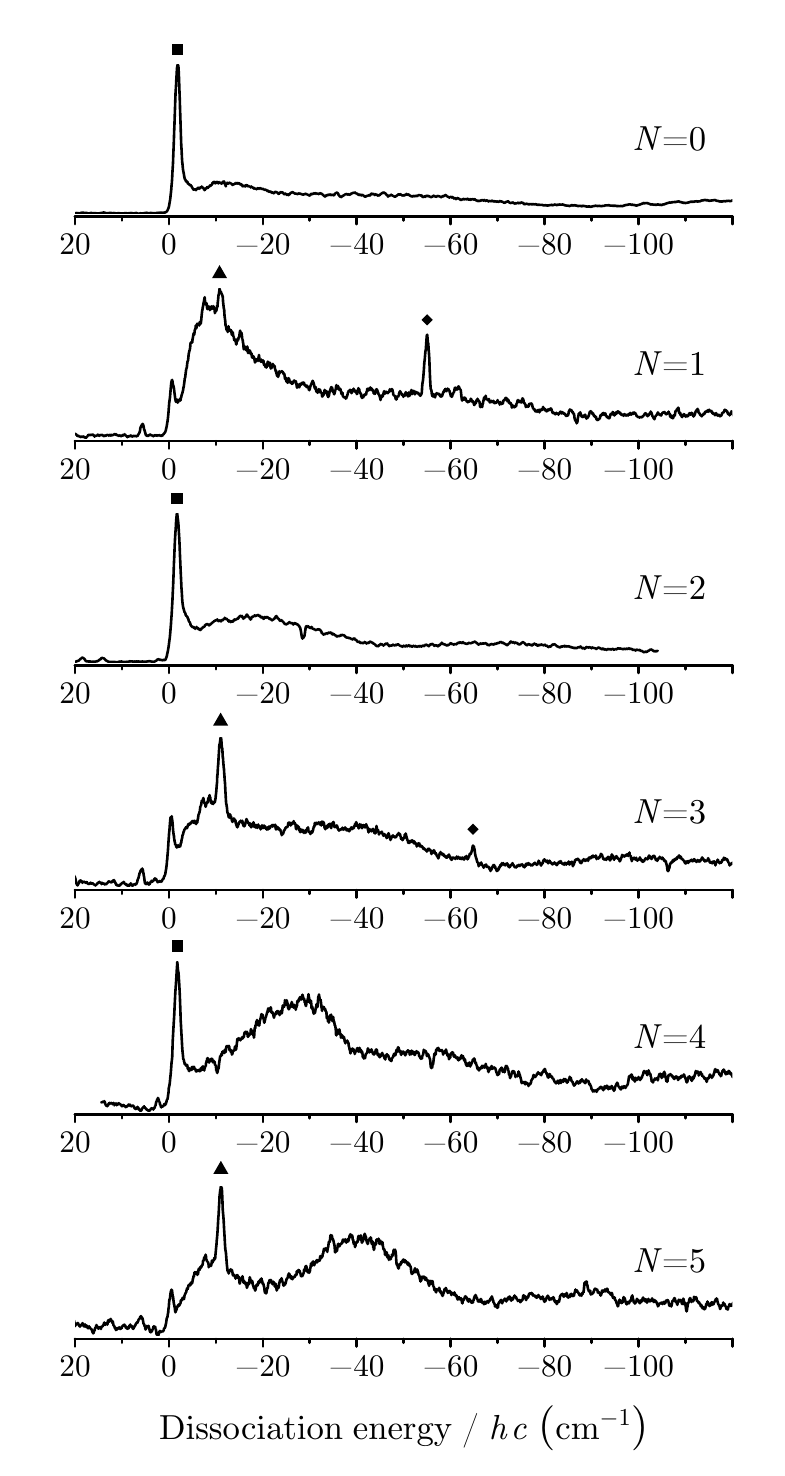}
\caption{PFI-ZEKE photoelectron spectra of H$_2$ recorded near the dissociative-ionization threshold from the $\bar{\rm H}\ (v=11,N=0-5)$ intermediate levels. The origin of the abscissa scale has been placed at the position of the dissociative-ionization threshold. The spectra have been corrected for the field-induced shift of the ionization thresholds, so that the negative dissociation energies correspond to the H$^+$ + H(1s) fragment kinetic energy. The lines marked by $\blacklozenge$ are transitions from the selected B intermediate state to the X$^+$(2,5) level (see text). The (17,7) and (18,4) resonances are marked by $\blacktriangle$ and $\blacksquare$, respectively. 
\label{fig2}}
\end{figure}

The resonant three-photon excitation sequence through the B and $\bar{\rm H}$ intermediate states used to access the dissociation threshold of H$_2^+$ makes it possible to study the dependence of the intensity distribution of the PFI-ZEKE photoelectron spectra on the selected rotational level of the $\bar{\rm H}(v=11)$ state. The distribution of population over the ground state rotational levels in the supersonic beam is dominated by the contributions from the $N=0$ (para H$_2$) and $N=1$ (ortho H$_2$) levels. Using P- and R-branch lines for the B-X and $\bar{\rm H}$-B transitions, rotational levels of the $\bar{\rm H}$ state with $N$ in the range 0--3 can be reached. The $N=2$ and 3 ground-state rotational levels could also be observed when the excitation laser pulse was triggered so as to probe either the very early or the late part of the H$_2$ gas pulses. From these levels and exploiting R-branch transitions for both electronic transitions, $\bar{\rm H}$ rotational levels with $N$ up to 5 could be accessed. Fig.~\ref{fig2} compares the PFI-ZEKE photoelectron spectra of H$_2$ recorded in the region of the H$^+$ + H(1s) dissociation threshold from the $N=0-5$ rotational levels of the $\bar{\rm H}$ state. The spectra recorded from the $N=0,\ 2$ and 4 levels reveal as most prominent spectral feature the transition to the X$^+$ ($v^+=18,N^+=4$) shape resonance located just above the dissociation threshold. The spectra recorded from the $N=1,\ 3$ and 5 are dominated by the transition to the X$^+$ ($v^+=17,N^+=7$) shape resonance. 

In addition to these shape resonances, the PFI-ZEKE photoelectron signal in the continuum displays slow variations with the excitation energy and which depend on the selected $\bar{\rm H}$-state rotational level.
The $\bar{\rm H}(v^+=11)$ state is sufficiently strongly bound that the centrifugal-potential contribution of states with $N=0-5$ does not affect the vibrational wave function (see Fig.~\ref{fig4} and discussion in Section~\ref{theory}). In contrast, the centrifugal-potential term has a strong influence on the vibrational wavefunctions of the bound and quasi-bound levels of H$_2^+$ near the dissociation threshold. The increasing $N$ value of the selected $\bar{\rm H}(v=11)$ levels is mirrored by an increase in the average value of the rotational quantum number of the ionic levels produced by photoionization. We therefore attribute the slow fluctuations of the photoelectron signal above the H$^+$~+~H(1s) dissociation threshold oberserved in Fig.~\ref{fig2} to the variations in the Franck-Condon factors, arising from the energy and $N^+$ dependences of the X$^+$ and A$^+$ vibrational wave functions. Support for this interpretation will be presented in Section~\ref{res:FC}.

\renewcommand{\thefootnote}{\fnsymbol{footnote}}
\begin{longtable}{llrrr}
%\resizebox{\textwidth}{!}{
%\begin{tabular}{lllrrrrr}
\caption{Measured level positions of para and ortho H$_2^+$ with respect to the X$^+$(17,6) and X$^+$(18,3) levels respectively (Exp-A) and corresponding dissociation energies (Exp-B). The last column lists the dissociation energies given by Moss \cite{moss93a}. All values are in cm$^{-1}$ .\label{tab:1}}\\
\toprule
$v^+$ & $N^+$ &                 Exp-A &                 Exp-B &                 Calculated \cite{moss93a} \\\midrule\endfirsthead
$v^+$ & $N^+$ &                 Exp-A &                 Exp-B &                 Calculated \cite{moss93a} \\\midrule\endhead
  2 &   1 &      -645.65(5) &     17072.15(5) &   17072.1015  \\
  2 &   2 &      -495.85(8) &     16968.10(8) &   16968.1122  \\
  2 &   3 &      -386.73(4) &     16813.23(4) &   16813.1836  \\
  2 &   4 &      -136.29(3) &     16608.54(3) &   16608.5514  \\
  2 &   5 &        70.64(5) &     16355.86(5) &   16355.8142  \\
 14 &   6 &     -1005.78(5) &      1024.35(5) &    1024.3432  \\
 14 &   7 &      -903.20(3) &       909.23(3) &     909.2448  \\
 14 &   8 &      -764.98(3) &       783.55(3) &     783.5210  \\
 14 &   9 &     -643.369(26) &      649.402(26) &     649.4257  \\
 15 &   0 &      -830.17(3) &       848.74(3) &     848.7019  \\
 15 &   1 &     -827.112(23) &      833.144(23) &     833.1619  \\
 15 &   2 &     -783.836(27) &      802.407(27) &     802.4019  \\
 15 &   3 &     -751.018(26) &      757.051(26) &     757.0589  \\
 15 &   4 &     -679.526(26) &      698.096(26) &     698.0824  \\
 15 &   5 &     -620.688(19) &      626.721(19) &     626.7297  \\
 15 &   6 &     -525.971(29) &      544.541(29) &     544.5638  \\
 15 &   7 &     -447.439(25) &      453.472(25) &     453.4604  \\
 15 &   8 &      -337.07(3) &       355.64(3) &     355.6303  \\
 15 &   9 &     -247.642(29) &      253.675(29) &     253.6735  \\
 16 &   0 &     -413.309(27) &      431.880(27) &     431.8657  \\
 16 &   1 &     -413.901(22) &      419.934(22) &     419.9309  \\
 16 &   2 &     -377.845(27) &      396.416(27) &     396.4046  \\
 16 &   3 &     -355.948(20) &      361.981(20) &     361.9742  \\
 16 &   4 &     -299.122(25) &      317.692(25) &     317.6752  \\
 16 &   5 &     -258.875(22) &      264.908(22) &     264.9021  \\
 16 &   6 &      -186.87(3) &       205.44(3) &     205.4345  \\
 16 &   7 &     -135.462(29) &      141.495(29) &     141.4972  \\
 16 &   8 &       -57.36(5) &        75.93(5) &      75.8998  \\
 16 &   9 &        -6.39(6) &        12.42(6) &      12.4157  \\
 17 &   0 &      -137.02(3) &       155.59(3) &     155.6233  \\
 17 &   1 &     -141.698(25) &      147.731(25) &     147.7104  \\
 17 &   2 &     -113.693(29) &      132.264(29) &     132.2789  \\
 17 &   3 &     -104.108(23) &      110.141(23) &     110.1302  \\
 17 &   4 &      -63.936(28) &       82.506(28) &      82.5046  \\
 17 &   5 &      -45.145(23) &       51.178(23) &      51.1587  \\
 17 &   6 &        0.000\footnotemark[1] &       18.571( 0) &      18.5707  \\
 17 &   7 &        17.11(6) &       -11.08(6) & 		$\cdots$  \\
 17 &   7 &        $\varGamma=0.56(8)$\footnotemark[2] &        & 		$\cdots$  \\
 18 &   0 &        -5.46(11) &        24.03(11) &      24.0435  \\
 18 &   1 &       -14.56(4) &        20.59(4) &      20.6115  \\
 18 &   2 &         4.349(27) &        14.222(27) &      14.2484  \\
 18 &   3 &        0.000\footnotemark[1] &        6.033( 0) &       6.0329  \\
 18 &   4 &        20.41(4) &        -1.84(4) & 		$\cdots $  \\
  18 &   4 &        $\varGamma=0.21(7)$\footnotemark[2] &         & 		$\cdots $  \\
  0 &   0 &         2.61(3) &         3.42(3) &   3.4373  \\
  0 &   1 &        16.08(4) &         2.49(4) &   2.5080  \\
  0 &   2 &        5.265(16) &        0.768(16) &   0.8081  \\
  \bottomrule
  \footnotetext[1]{Reference levels (see text).}
  \footnotetext[2]{From Ref. \cite{beyer16a}.}
\end{longtable}

%
% T H E O R Y
%
\section{Theoretical considerations and computational details}
\label{theory}

In this section, we briefly present the procedure followed in the present work to calculate the bound and quasibound rovibronic states of H$_2^+$. The Schrödinger equation for the nuclei is solved at different levels of approximation \cite{carrington84a,leach95a} to determine the level energies and line widths characterizing the spectra and dynamics of H$_2^+$. 

We start from the complete non-relativistic Hamiltonian for the internal motion of H$_2^+$, which may be written in atomic units as \cite{teller70a,carrington84a}
\begin{equation}
\label{eq:fullHam}
\mathcal{H_\text{int}} = -\frac{\nabla^2_r}{2} - \frac{1}{r_{\text{A}}} - \frac{1}{r_{\text{B}}} + \frac{1}{R} -\frac{\nabla^2_R}{2\mu} -\frac{\nabla^2_r}{8\mu},
\end{equation}   
where $\mu$ is the reduced mass of the nuclei and $\mathbf{r}$ and $\mathbf{R}=\mathbf{R}_\text{A}-\mathbf{R}_\text{B}$ are the position of the electron relative to the geometric center and the relative position of the nuclei A and B, respectively. 

\begin{figure}[ht]
\centering
\includegraphics[width=0.6\columnwidth]{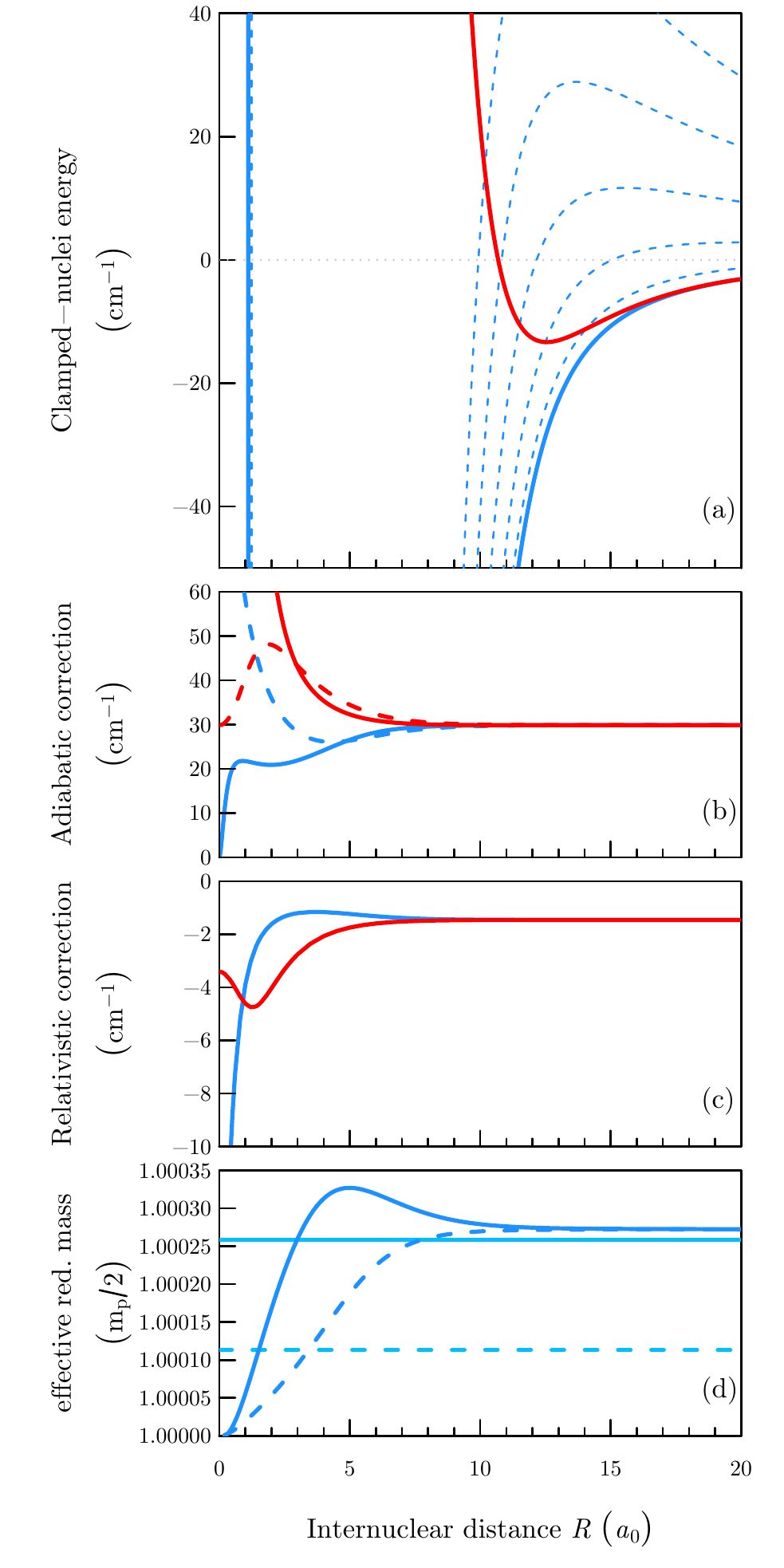}
\caption{(a) Clamped-nuclei (Born-Oppenheimer) potential energy functions of the X$^+$ $^2\Sigma_g^+$ and A$^+$ $^2\Sigma_u^+$ states of H$_2^+$ calculated in this work with respect to the dissociation energy (gray dotted line). The blue dashed curves include the centrifugal potential term for $N^+=2,4,6,8$ and 10. (b) $H_1$ (full lines) and $H_2$ (dashed lines) adiabatic corrections calculated in this work. (c) Relativistic corrections \cite{howells90a}. (d) Effective reduced-mass functions for the vibrational (full lines) and rotational (dashed lines) motions used to evaluate the nonadiabatic corrections of the X$^+$ state of H$_2^+$ taken from the work of Jaquet and Kutzelnigg \cite{jaquet08a} (dark blue) and Moss \cite{moss96a} (light blue). In (a)-(c), the functions drawn in red and blue color are for the X$^+$ and A$^+$ states, respectively.
\label{fig3}}
\end{figure}

\subsection{Born-Oppenheimer Solutions}
The clamped-nuclei Hamiltonian is obtained from \eqref{eq:fullHam} by setting the last two terms to zero, which corresponds to assuming that the nuclei have an infinitely heavy mass. The resulting electronic Born-Oppenheimer equation is given by
\begin{equation}
\label{eq:BOeq}
\left( -\frac{\nabla_r^2}{2} - \frac{1}{r_{\text{A}}} - \frac{1}{r_{\text{B}}} + \frac{1}{R} \right)\psi_t(r;R) = U^\text{BO}(R)\psi_t(r;R).
\end{equation} 
This equation is separable in prolate spheroidal coordinates $(\xi,\eta,\gamma)$ and can be solved by variational methods \cite{bates53a,hunter66a,hunter67a}. The explicit expressions for all operators can be obtained from the general two-electron forms derived in the famous article of Kolos and Wolniewicz \cite{kolos63a} by replacing $\phi$ by 0 and $\Lambda/2$ by $\Lambda^+$, as defined in their work. 
The ansatz for the electronic wave function $\psi_t(r;R)$ used here is \cite{hunter67a}
\begin{equation}
\label{eq:eWF}
\psi_t(\xi,\eta,\gamma;R) = f(\xi;R)g(\eta;R)n(\gamma),
\end{equation}
with 
\begin{align}
g(\xi;R) &= (\xi^2-1)^{\Lambda^+/2} (\xi+1)^{(R/p)-\Lambda^+-1}\exp{(-p\xi)} \sum_{n=0}^{\infty} g_n(R) \left(\frac{\xi-1}{\xi+1}\right)^n, \\
f(\eta;R) &= \sum_{s=0}^{\infty} f_s(R) P^{\Lambda^+}_{\Lambda^++s}(\eta),~\text{and} \\
n(\gamma) &= \sqrt{\frac{1}{2\pi}}\exp{\left[ i\Lambda^+\gamma\right]},
\end{align}
where $P^{\Lambda^+}_{\Lambda^++s}(\eta)$ are associated Legendre polynomials and $p = R\sqrt{-(U^\text{BO}(R)-1/R)/2}$.
The expansion coefficients $g_n$ and $f_s$ were determined by requiring that the eigenvalue $A$ simultaneously satisfies
\begin{align}
\label{eq:EVproblem}
\mathbf{G}\cdot\mathbf{g} &= -A\mathbf{g} \notag\\
\mathbf{F}\cdot\mathbf{f} &= A\mathbf{f}, 
\end{align}
where the tridiagonal matrices $\mathbf{G}$ and $\mathbf{F}$ are given in terms of $R$, $p$ and $\Lambda^+$ \cite{hunter67a}.

\emph{Numerical details.} The calculation was started at $R=0$ using the known values of all terms occurring in \eqref{eq:EVproblem} in the united-atom limiting case \cite{hunter67a}. The solution was propagated up to $R=150~a_0$, assuming a smooth behaviour of $A$ and adjusting the size of the basis set at each internuclear distance $R$ so that the ratio of the first to the last expansion coefficient was at least $10^{15}$.
For $R>150~a_0$, $U^\text{BO}(R)$ was extrapolated using $U^\text{BO}(R)=-0.5-2.25R^{-4}-7.292R^{-6}-7506R^{-8}$, where we used the BO energy and the polarizability of the H atom \cite{coulson41a} and determined the two last coefficients through a linear fit to our ab initio values for $R>100~a_0$. For the proton-to-electron mass ratio we used $m_\text{p}=1836.15267389(17)$ and $E_\text{h}/hc=219474.6313702(13)$~cm$^{-1}$ \cite{mohr14a}.  
The BO potential energy functions of the $X^+$ (blue) and the $A^+$ (red) states are depicted in Fig. \ref{fig3}a and are given in Ref. \cite{beyer16b}.

\subsection{Adiabatic Solutions}
Using the Born expansion ansatz for the complete molecular wave function $\Psi_\text{int}=\frac{1}{R}\sum_t \psi_t(r;R)\phi_t(R)$ together with the full molecular Hamiltonian \eqref{eq:fullHam}, a system of coupled equations is obtained \cite{carrington84a}
\begin{align}
\label{eq:CE}
 &\left\{  - \frac{\nabla^{2}_{R}}{2\mu}  + U^\text{BO}(R)  + \braket{H'_1}(R) + \braket{H'_2}(R) \right\} \frac{\phi_s(R)}{R}  \notag\\
 &+ \sum_{s\neq t} \Biggr\{             \int \psi_s^*(r;R) \left[-\frac{\nabla^{2}_{R}}{2\mu} - \frac{\nabla^2_r}{8\mu}   \right]  \psi_t(r;R) \text{d}\tau \notag\\ 
&+  \psi_s^*(r;R) \left[-\frac{\nabla_{R}}{\mu} \right] \psi_t(r_1;R) \text{d}\tau\cdot \nabla_{R} \Biggr\} \frac{\phi_t(R)}{R} = E_\text{int} \frac{\phi_s(R)}{R}.
\end{align}
In Eq. \eqref{eq:CE}
\begin{align}
\braket{H'_1}(R) &= \int \psi_s^*(r;R) \left[ - \frac{\nabla^{2}_{R}}{2\mu} \right]  \psi_s(r;R)  \text{d}\tau\\
\braket{H'_2}(R) &= \int \psi_s^*(r;R) \left[ - \frac{\nabla^2_r}{8\mu} \right] \psi_s(r;R)  \text{d}\tau
\end{align}
are the adiabatic corrections. A representation of each electronic state by a single potential-energy curve is preserved in the adiabatic approximation by neglecting the sum in \eqref{eq:CE}, because the separation of the electronic and nuclear motions remains exact in this approximation. The adiabatic potential-energy curve is then given by $U^\text{ad}(R)=U^\text{BO}(R) + \braket{H'_1}(R) + \braket{H'_2}(R)$. The adiabatic corrections were calculated as explained in Ref. \cite{wolniewicz78a} and are given in Ref. \cite{beyer16b}.

\emph{Numerical details.} 
For $R>150~a_0$, the adiabatic corrections were extrapolated using $\braket{H'_1}(R)+\braket{H'_2}(R)=-1/(2m_\text{p})-0.003676R^{-4}+0.01169R^{-6}$, where we used the adiabatic correction of the H atom and determined the two last coefficients through a linear fit to our ab initio values for $R>100~a_0$. 
The two functions $\braket{H'_1}(R)$ and  $\braket{H'_2}(R)$ are displayed in Fig. \ref{fig3}b as full and dashed lines respectively.

\subsection{Nonadiabatic Solutions}
The nonadiabatic interactions correspond to the off-diagonal elements in \eqref{eq:CE}. Various methods exist for the calculation of nonadiabatic rovibrational energies, for which we refer the reader to the excellent review of Leach and Moss \cite{leach95a}.
We followed here the approach consisting of introducing $R$-dependent reduced masses, which allows one to remove the off-diagonal coupling elements in \eqref{eq:CE} and so retain the concept of a single electronic potential curve. This idea, worked out in detail by Moss \cite{moss93a}, is particularly useful for the electronic ground state, because it is usually well separated from excited states. It was successfully applied to H$_2^+$ by Kutzelnigg and Jaquet \cite{kutzelnigg07a, jaquet08a} among others.
For the $X^+~^2\Sigma_g^+$ electronic ground state, we used the $R$-dependent vibrational and rotational reduced masses calculated in \cite{jaquet08a} using a LCAO ansatz for the electronic wave function and taking into account the leading nonadiabatic correction term resulting from the interaction of the electronic ground state with the 2s$\sigma_\text{g}$, 2p$\sigma_\text{g}$ and 2p$\pi_\text{g}$ states.
For the $A^+~^2\Sigma_u^+$ first excited state, which has a shallow minimum at ~12$a_0$, we used the atomic reduced mass to account for the leading-order nonadiabatic corrections. This is known to be a good approximation especially at large internuclear distances, as can be seen in Fig. \ref{fig3} for the electronic ground state.   

\emph{Numerical details.} Vibrational and rotational masses $\mu_\text{vib}^{-1}=\mu^{-1}\left(1+A_\mu(R)/m_\text{p} \right)$ and $\mu_\text{rot}^{-1}=\mu^{-1}\left(1+B_{\mu,\text{pol}}(R)/m_\text{p} \right)$ were determined using $A_\mu(R)$ and $B_{\mu,\text{pol}}(R)$ as given in \cite{jaquet08a}. The values were interpolated using a cubic spline and extrapolated for $R>20~a_0$ using $A_\mu(R)=-1/2 - 60.60R^{-4}$ and $B_{\mu,\text{pol}}(R)=-1/2 + 28.14R^{-4}$.
Fig. \ref{fig3}d presents the $R$-dependent reduced-mass functions for the $X^+$ state.

\subsection{Relativistic and radiative corrections}
The relativistic corrections $\braket{H_\text{rel}}(R)$ can be obtained in a straightforward way using the solutions of Eq. \eqref{eq:BOeq} as described in \cite{howells90a}. The calculation of the radiative corrections is more involved and the $R$ dependence is only known for rather small internuclear distances. Moss \cite{moss93a} used the values calculated by Bukowski et al. \cite{bukowski92a} and extrapolated them to larger internuclear distances using the electron density. He reported the final radiative corrections for individual rovibrational levels.

\emph{Numerical details.}  We used the relativistic corrections given in \cite{howells90a} for the $X^+$ and $A^+$ state and interpolated them using a cubic spline. Because the $R$ dependence of the radiative corrections used in \cite{moss93a} was not given explicitly by Moss, we added his corrections to our level energies.

\subsection{Calculation of bound states}
Bound-state energies were obtained by numerical integration of Eq. \eqref{eq:CE}, which reads in spherical polar coordinates as
\begin{equation}
\label{eq:SEN}
\left[-\frac{1}{2\mu_\text{vib}}\frac{\text{d}^2}{\text{d}R^2} + U + \frac{N^+(N^++1)}{2\mu_\text{rot}R^2} - E_i\right]\phi_i=0, 
\end{equation}
using the renormalized Numerov method as introduced by Johnson \cite{johnson77a}. The influence of the centrifugal contribution on the $R$-dependent potential (third term) can bee seen in Fig. \ref{fig3}a. Within the various levels of approximation, we used (BO = Born-Oppenheimer, AD = adiabatic, NA = nonadiabatic approximation)
\begin{align}
\text{BO:} \quad& U= U^\text{BO}, \mu_\text{vib}=\mu_\text{rot}=\frac{m_\text{p}}{2}, \label{eq:UAD}\\
\text{AD:} \quad& U= U^\text{BO}+\braket{H'_1}+\braket{H'_2},~\mu_\text{vib}=\mu_\text{rot}=\frac{m_\text{p}}{2},\\
X^+~\text{NA:} \quad& U= U^\text{BO}+\braket{H'_1}+\braket{H'_2},  \notag\\
	& \mu_\text{vib}^{-1}=\mu^{-1}\left(1+A_\mu/m_\text{p} \right),~\mu_\text{rot}^{-1}=\mu^{-1}\left(1+B_{\mu,\text{pol}}/m_\text{p} \right), \\ 
A^+~\text{NA:} \quad& U= U^\text{BO}+\braket{H'_1}+\braket{H'_2},~\mu_\text{vib}=\mu_\text{rot}=\frac{m_\text{p}(m_\text{p}+1)}{2m_\text{p}}.
\end{align}
Relativistic energies were obtained by including $\braket{H_\text{rel}}$ in $U$.
We used five different step sizes $h=0.01,0.02,0.03,0.04$ and $0.05~a_0$ on the interval $[0.01,200]~a_0$ and relied on the Richardson extrapolation $E(h)=E(h=0)+c_1h^4+c_2h^6$ to obtain the level energies for $h\to0$.
Dissociation energies were obtained by subtracting the corresponding asymptotic energy of the hydrogen atom from the calculated level energies using
\begin{align}
E_\text{H}^{(\text{BO})} &= -1/2~\text{a.u.} \\
E_\text{H}^{(\text{AD})} &= \frac{1-m_\text{p}}{2m_\text{p}}~\text{a.u.}\\
E_\text{H}^{(\text{NA})} &=  -\frac{1}{2(1+m_\text{p}^{-1}) }~\text{a.u.}\label{eq:E_infty_NA}\\
E_\text{H}^{(\text{NA,rel,rad})} &= \left[ -\frac{1}{2(1+m_\text{p}^{-1}) }~+~\frac{(-1.46092+0.27066)~\text{cm}^{-1}}{2\mathcal{R_\infty}}\right]~\text{a.u.}~,
\end{align}
where $\mathcal{R_\infty}$ is the Rydberg constant in cm$^{-1}$. 

\subsection{Calculation of resonances}\label{ssec:res}
For the localization of resonances, we integrated Eq. \eqref{eq:SEN} on a coarse energy grid outwards to the outermost turning point and counted the number of nodes. At the energies at which a new node appeared, we integrated the wave function in the vicinity of these energies to large $R$ and matched the wave function with its asymptotic form
\begin{equation}
\lim_{R\to\infty}\phi(R;k) = A_kkR \left(j_{N^+}(kR)\cos\delta_{N^+} - n_{N^+}(kR)\sin\delta_{N^+} \right),  
\end{equation} 
where $j_{N^+}$ and $n_{N^+}$ are the spherical Bessel functions and $k=\sqrt{2\mu(E-U)}$. Energy-normalized continuum wave functions are obtained by scaling the amplitude to \cite{fano86a}
\begin{equation}\label{eq:ENorm}
A_k = \sqrt{\frac{2\mu}{\pi k}}.
\end{equation}
The phase shift for a given energy $\delta_{N^+}(E)$ was obtained by using the values of the wave function at the two outermost grid points $R_\text{a}$ and $R_\text{b}=R_\text{max}$ using
\begin{equation}
\tan{\delta_{N^+}} = \frac{ Kj_{N^+}(kR_\text{a}) - j_{N^+}(kR_\text{b}) }{ Kn_{N^+}(kR_\text{a}) - n_{N^+}(kR_\text{b}) }; ~ K=\frac{ R_\text{a}\phi_{N^+}(R_\text{b}) }{ R_\text{b}\phi_{N^+}(R_\text{a})  }.
\end{equation}
The energy grid in the vicinity of a resonance was made adaptive by requiring a certain number of points per phase jump $\pi$.

The position $E_\text{res}$ and the full width at half maximum $\varGamma^{\text{(BW)}}$ were determined in a nonlinear least-squares fit of a Breit-Wigner-like formula \cite{smith71a}
\begin{equation}\label{eq:BW}
\delta_{N^+}(E) = \sum_{j=0}^{2} \delta_{N^+}^{(j)}E^j + \arctan\left[\frac{\varGamma^{\text{(BW)}}/2}{E_\text{res}-E} \right] 
\end{equation} 
using an energy-dependent background phase shift.
Especially for broad resonances located near the H$^+$ + H(1s) dissociation threshold, the parametrization in Eq. \eqref{eq:BW} becomes ambiguous and one usually determines the resonance parameters within the collision-time-delay approach developed by Smith \cite{smith60a}. For a single channel, the scattering matrix is $S=\exp\left[ 2i\delta_{N^+} \right]$ and the lifetime matrix is given by
\begin{equation}
Q=-iS^*\frac{dS}{dE}= 2 \frac{d\delta_{N^+}}{dE}.
\end{equation}  
The resonance position corresponds to the position of the maximum of $Q$, i.e. the energy at which $\frac{d\delta_{N^+}}{dE}$ is maximal. The level width is given by
\begin{equation}\label{eq:Qgamma}
\varGamma^\text{(Q)} = \frac{4}{Q(E_\text{res})}= \frac{2}{\frac{d\delta_{N^+}}{dE}\big|_{E=E_\text{res}}}.
\end{equation}
The derivatives of the phase shift $\delta_{N^+}$ with respect to the energy were evaluated numerically by interpolating $\delta_{N^+}$ using a cubic spline and taking the analytic derivative of the spline. 
By comparing Eqs. \eqref{eq:BW} and \eqref{eq:Qgamma} one obtains
\begin{equation}\label{eq:BWQ}
\varGamma^\text{(Q)} \simeq \varGamma^\text{(BW)}\frac{2}{2+\varGamma^\text{(BW)}\delta_{N^+}^{(1)} + 2\varGamma^\text{(BW)}\delta_{N^+}^{(2)}E_\text{res}},
\end{equation}
so that $\varGamma^\text{(Q)}$ is smaller than $\varGamma^\text{(BW)}$, particulary for broad resonances. 

\begin{landscape}
\footnotesize
\begin{longtable}{lllrrrrrr}
%\resizebox{\textwidth}{!}{
%\begin{tabular}{lllrrrrr}
\caption{Positions and widths of the ninteen shape resonances of H$_2^+$ not reported by Moss, calculated at different levels of approximation: Born-Oppenheimer (BO), adiabatic (AD), Moss masses (MM) \cite{moss96a},  nonadiabatic (NA), nonadiabatic with relativistic corrections (NArel), nonadiabatic with relativistic and radiative corrections (NArelrad). The values listed for the widths correspond to $\varGamma^\text{BW}$ and $\varGamma^\text{BW}-\varGamma^\text{Q}$ is given in parentheses. The last three entries of the table give resonance parameters for three narrow shape resonances and compare with the positions calculated by Moss \cite{moss93a}.\label{tab:2}}\\
\toprule
$v^+$ & $N^+$ &                      &           BO &           AD &           MM &           NA &        NArel &     NArelrad \\\midrule\endfirsthead
$v^+$ & $N^+$ &                      &           BO &           AD &           MM &           NA &        NArel &     NArelrad \\\midrule\endhead
 18 &   4 &       $E_\text{res}$ &       -1.991 &       -1.879 &       -1.877 &       -1.876 &       -1.879 &       -1.878 \\
    &     &          $\varGamma$ &        0.259~( 0.002) &        0.206~( 0.002) &        0.193~( 0.001) &        0.193~( 0.001) &        0.194~( 0.001) &  \\
 17 &   7 &       $E_\text{res}$ &      -11.699 &      -11.210 &      -11.099 &      -11.094 &      -11.104 &      -11.104 \\
    &     &          $\varGamma$ &        0.232~( 0.001) &        0.177~( 0.001) &        0.163~( 0.001) &        0.162~( 0.001) &        0.163~( 0.001) &  \\
 16 &  10 &       $E_\text{res}$ &      -43.219 &      -42.453 &      -42.269 &      -42.255 &      -42.273 &      -42.272 \\
    &     &          $\varGamma$ &        1.108~( 0.008) &        0.961~( 0.006) &        0.921~( 0.005) &        0.921~( 0.005) &        0.924~( 0.005) &  \\
 15 &  13 &       $E_\text{res}$ &     -113.545 &     -112.727 &     -112.543 &     -112.514 &     -112.534 &     -112.532 \\
    &     &          $\varGamma$ &        9.185~( 0.210) &        8.638~( 0.193) &        8.492~( 0.202) &        8.488~( 0.201) &        8.501~( 0.203) &  \\
 14 &  15 &       $E_\text{res}$ &     -150.399 &     -149.090 &     -148.783 &     -148.742 &     -148.775 &     -148.773 \\
    &     &          $\varGamma$ &        2.836~( 0.027) &        2.571~( 0.023) &        2.503~( 0.021) &        2.502~( 0.022) &        2.508~( 0.022) &  \\
 13 &  17 &       $E_\text{res}$ &     -199.990 &     -198.210 &     -197.797 &     -197.739 &     -197.786 &     -197.783 \\
    &     &          $\varGamma$ &        1.100~( 0.004) &        0.972~( 0.003) &        0.941~( 0.003) &        0.941~( 0.003) &        0.944~( 0.003) &  \\
 12 &  19 &       $E_\text{res}$ &     -269.686 &     -267.509 &     -267.017 &     -266.940 &     -266.999 &     -266.995 \\
    &     &          $\varGamma$ &        0.770~( 0.001) &        0.678~( 0.001) &        0.657~( 0.001) &        0.656~( 0.001) &        0.658~( 0.001) &  \\
 11 &  21 &       $E_\text{res}$ &     -364.895 &     -362.399 &     -361.855 &     -361.755 &     -361.825 &     -361.820 \\
    &     &          $\varGamma$ &        0.973~( 0.002) &        0.864~( 0.002) &        0.839~( 0.002) &        0.839~( 0.002) &        0.841~( 0.002) &  \\
 10 &  23 &       $E_\text{res}$ &     -489.019 &     -486.275 &     -485.699 &     -485.573 &     -485.653 &     -485.646 \\
    &     &          $\varGamma$ &        1.827~( 0.008) &        1.648~( 0.006) &        1.609~( 0.006) &        1.608~( 0.006) &        1.613~( 0.006) &  \\
  9 &  25 &       $E_\text{res}$ &     -644.015 &     -641.083 &     -640.495 &     -640.338 &     -640.425 &     -640.418 \\
    &     &          $\varGamma$ &        4.151~( 0.042) &        3.811~( 0.035) &        3.738~( 0.034) &        3.736~( 0.034) &        3.745~( 0.034) &  \\
  8 &  27 &       $E_\text{res}$ &     -831.031 &     -827.953 &     -827.361 &     -827.169 &     -827.263 &     -827.254 \\
    &     &          $\varGamma$ &        9.713~( 0.235) &        9.069~( 0.204) &        8.936~( 0.198) &        8.933~( 0.198) &        8.951~( 0.198) &  \\
  7 &  29 &       $E_\text{res}$ &    -1051.160 &    -1047.934 &    -1047.339 &    -1047.107 &    -1047.207 &    -1047.197 \\
    &     &          $\varGamma$ &       21.242~( 1.090) &       20.123~( 0.981) &       19.902~( 0.960) &       19.898~( 0.960) &       19.930~( 0.963) &  \\
  6 &  30 &       $E_\text{res}$ &    -1023.942 &    -1019.316 &    -1018.474 &    -1018.236 &    -1018.379 &    -1018.366 \\
    &     &          $\varGamma$ &        0.521~( 0.000) &        0.465~( 0.000) &        0.454~( 0.000) &        0.454~( 0.000) &        0.456~( 0.000) &  \\
  5 &  32 &       $E_\text{res}$ &    -1306.304 &    -1301.540 &    -1300.710 &    -1300.425 &    -1300.575 &    -1300.560 \\
    &     &          $\varGamma$ &        2.451~( 0.012) &        2.236~( 0.010) &        2.197~( 0.010) &        2.197~( 0.010) &        2.204~( 0.010) &  \\
  4 &  34 &       $E_\text{res}$ &    -1626.916 &    -1622.011 &    -1621.190 &    -1620.852 &    -1621.009 &    -1620.993 \\
    &     &          $\varGamma$ &        8.140~( 0.147) &        7.577~( 0.126) &        7.480~( 0.123) &        7.482~( 0.123) &        7.499~( 0.124) &  \\
  3 &  36 &       $E_\text{res}$ &    -1988.537 &    -1983.404 &    -1982.577 &    -1982.180 &    -1982.345 &    -1982.328 \\
    &     &          $\varGamma$ &       19.569~( 0.875) &       18.494~( 0.780) &       18.317~( 0.765) &       18.324~( 0.766) &       18.356~( 0.768) &  \\
  2 &  38 &       $E_\text{res}$ &    -2396.207 &    -2390.696 &    -2389.838 &    -2389.371 &    -2389.551 &    -2389.530 \\
    &     &          $\varGamma$ &       35.494~( 2.853) &       33.895~( 2.608) &       33.643~( 2.572) &       33.663~( 2.576) &       33.711~( 2.583) &  \\
  1 &  39 &       $E_\text{res}$ &    -2439.189 &    -2432.093 &    -2431.026 &    -2430.575 &    -2430.803 &    -2430.780 \\
    &     &          $\varGamma$ &        1.591~( 0.005) &        1.456~( 0.004) &        1.436~( 0.004) &        1.438~( 0.004) &        1.442~( 0.004) &  \\
  0 &  41 &       $E_\text{res}$ &    -2933.728 &    -2926.023 &    -2924.904 &    -2924.387 &    -2924.637 &    -2924.610 \\
    &     &          $\varGamma$ &        2.169~( 0.009) &        2.003~( 0.008) &        1.981~( 0.008) &        1.984~( 0.008) &        1.989~( 0.008) &  \\\hline
 15 &  12 &       $E_\text{res}$ &      -42.798 &      -41.330 &      -40.960 &      -40.938 &      -40.972 &      -40.971 \\
    &     &		$E_\text{res}$ \cite{moss93a} &   &        &        &        &        &      -40.9(1) \\
    &     &          $\varGamma$ \footnotemark[1] &     266E-05 &    179E-05 &    159E-05 &     159E-05 &    160E-05 & \\
 10 &  22 &       $E_\text{res}$ &     -265.907 &     -262.270 &     -261.483 &     -261.379 &     -261.480 &     -261.474 \\
     &     &		$E_\text{res}$ \cite{moss93a} &   &        &        &        &        &      -261.4826 \\
    &     &          $\varGamma$ \footnotemark[1]&   5.00E-05 &   3.76E-05 &   3.52E-05 &   3.51E-05 &   3.54E-05 & \\
  0 &  40 &       $E_\text{res}$ &    -2428.929 &    -2420.473 &    -2419.257 &    -2418.827 &    -2419.097 &    -2419.070 \\
      &     &		$E_\text{res}$ \cite{moss93a} &   &        &        &        &        &      -2419.23(2) \\
    &     &          $\varGamma$ \footnotemark[1]&    782E-05 &    692E-05 &    679E-05 &     681E-05 &     684E-05 & \\
    \bottomrule
%\end{tabular}
%}
\end{longtable}
\parbox[t]{\linewidth}{\raggedright\footnotemark[1]{ $\varGamma^\text{BW}$ and $\varGamma^\text{Q}$ are equal within the accuracy of our calculation.     } }
\end{landscape}

\begin{figure}[bt]
\centering
\includegraphics[width=1\columnwidth]{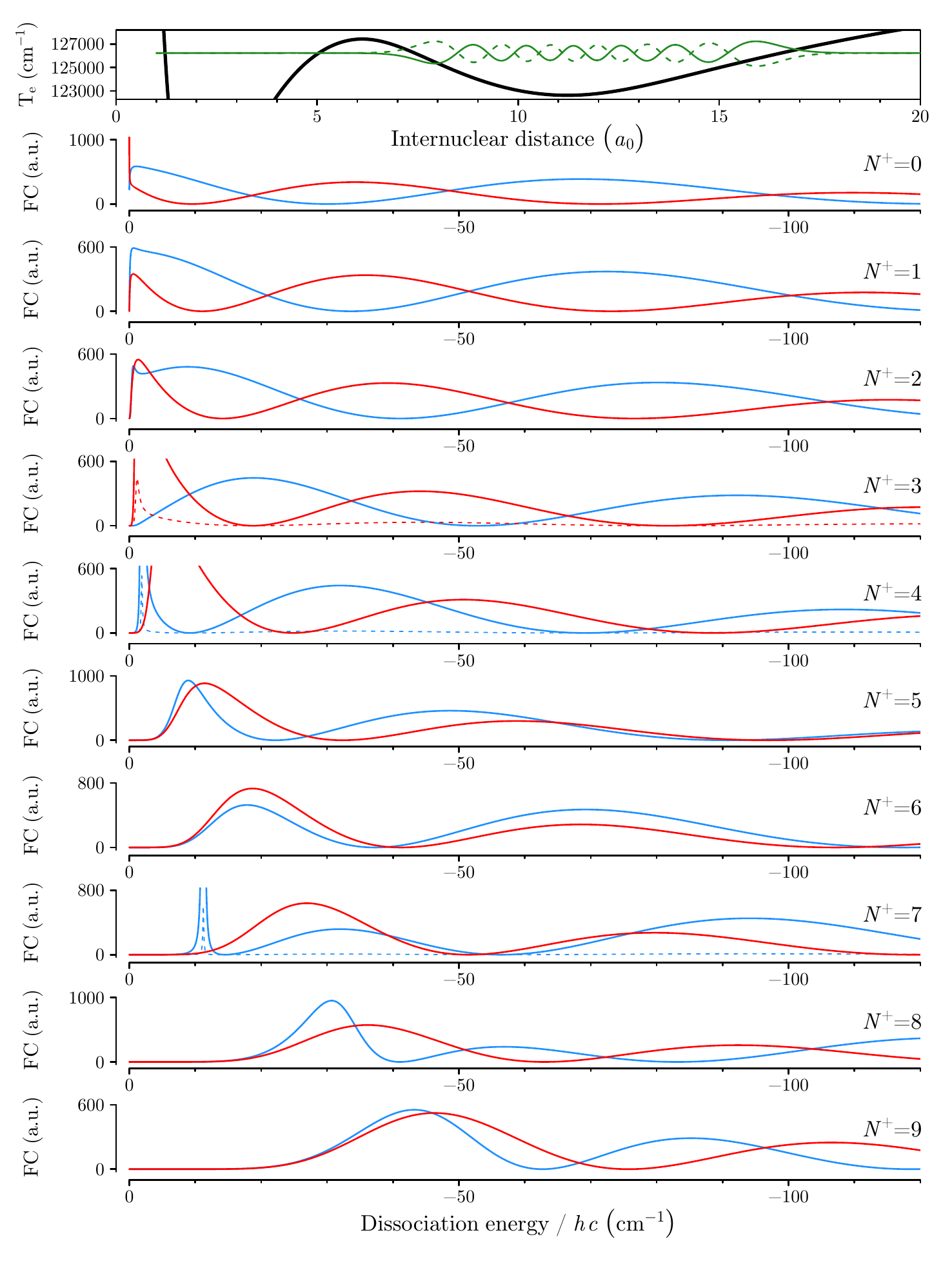}
\caption{Top panel: Vibrational wavefunctions obtained for the $N=0$ (solid) and $N=5$ (dashed) levels of the $\bar{\rm H}\ (v=11)$ state of H$_2$. Lower panels: Bound-continuum Franck-Condon factors for transitions from the $\bar{\rm H}\ (v=11)$ state to the H$^+$ + H dissociation chanels with angular momentum quantum numbers $N^+=0$--9, respectively. The solid blue and red lines corresponds to the X$^+$ and A$^+$ channels, respectively. The dashed lines presented on a reduced vertical scale demonstrate the dominance of the resonances over the continuum contributions.
\label{fig4}}
\end{figure}

\section{Computational results and discussion}

\subsection{Resonance positions and widths}

With the procedure outlined in Section \ref{theory}, the accuracy of the calculated level positions is limited by the approximative treatment of the nonadiabatic corrections through $R$-dependent reduced masses. As was shown by Moss \cite{moss96a}, the nonadiabatic correction to the dissociation energies is largest around $v^+=11$ and decreases for lower and higher vibrational quantum numbers. Taking into account the nonadiabatic effects using the $R$-dependent reduced masses of Jaquet and Kutzelnigg \cite{jaquet08a}, we verified for the observed bound states (see Tab. \ref{tab:1}), that the deviations between our results and the exact positions of Moss \cite{moss93a} are always less than 2~GHz. The evolution of these deviations is not smooth with $v^+$ but show maxima around $v^+=3$ and $v^+=15$ and minima around $v^+=0,8$ and 19. The effects of an approximate treatment of nonadiabatic effects using an $R$-dependent rotational reduced mass are amplified at high $N^+$ values. To quantify this effect, we recalculated three very narrow resonances, X$^+$ (0,40), (10,22) and (15,12). The largest deviation from the results of Moss \cite{moss93a} was observed for X$^+$ (0,40) level, but is less than 5~GHz, which we think is the upper limit for the remaining nonadiabatic corrections not included in a treatment based on $R$-dependent reduced masses.

The upper part of Table \ref{tab:2} provides a complete overview of the resonance positions and widths calculated at different levels of approximation. As expected, the dominant correction is the adiabatic one, followed by the nonadiabatic one, the relativistic corrections being typically less than 1~GHz for the level positions and the radiative corrections typically ten times less. Comparison of all data in Table \ref{tab:2} further indicates that the radiative corrections have a negligible effect (i. e., less than 1\textperthousand) on the resonance widths. 

Using the R-independent effective reduced masses determined by Moss in a least-square fit procedure to reproduce $X^+$ levels he calculated \cite{moss96a}, rather than the $R$-dependent ones, does reproduce the resonance positions within 10~GHz. This can be seen by comparing the columns labeled MM and NA in Table \ref{tab:2}. 

Our calculated positions for the (17,7) and (18,4) resonances agree with our experimental results within the experimental accuracy. The BO treatment of Davis and Thorson \cite{davis78a} led to values of $-2.0$~cm$^{-1}$, $-11.7$~cm$^{-1}$ for the positions and 0.29~cm$^{-1}$, 0.44~cm$^{-1}$ for the widths of the (17,7) and (18,4) resonances. It is thus necessary to include adiabatic and nonadiabatic effects to reproduce the experimental results.

The two methods we used to calculate the resonance widths (Eqs. \eqref{eq:BW} and \eqref{eq:Qgamma}) give identical results for narrow resonances because the energy dependence of the background phase shift is usually small and the quotient in Eq. \eqref{eq:BWQ} approaches 1. For broad resonances, the differences are significant and approach 10\% for the broadest one X$^+$ (2,38).
These differences are linked to the inherent difficulties one encounters when describing broad resonances with only their positions and their widths \cite{roy71a}.

Whereas the width we calculated for the (18,4) resonance agrees with the observed width (see Table \ref{tab:1} and Ref. \cite{beyer16a}), the width observed for the (17,7) resonance is almost four times as large as the calculated width. We have no explanation for this discrepancy. 
It is conceivable that the accuracy of the experimental determination of resonance widths by PFI-ZEKE photoelectron spectroscopy decreases with increasing fragment kinetic energy. Further work to understand the reason for this discrepancy is currently underway.

\subsection{Bound-continuum Franck-Condon factors}\label{res:FC}
To qualitatively account for the broad oscillations of the dissociative-ionization cross section observed in Fig. \ref{fig2}, we used a simplified treatment based on the calculation of bound-continuum Franck-Condon factors \cite{tellinghuisen85a} of the type
\begin{equation}
\text{FC} = \left| \int_0^\infty \phi_{\bar{H}(v=11)}(R)~ \phi^{(N^+)}_{X^+/A^+}(E,R) ~\text{d}R  \right|^2.
\end{equation}
The function $\phi_{\bar{H}(v=11)}(R)$ was obtained by numerical integration of the $H\bar{H}$ adiabatic potential \cite{wolniewicz98b} according to Eqs. \eqref{eq:SEN} and \eqref{eq:UAD}, also including relativistic corrections \cite{wolniewicz98a}. The centrifugal term does not significantly affect the nuclear wave function at the large internuclear distance characteristic for the $\bar{H}$ outer well (i. e., $R\ge 7~a_0$). This can be seen in the top panel of Fig. \ref{fig4} by comparison of the vibrational wave functions obtained for $v=11,N=0$ and $N=5$, depicted as red and green curves, respectively, and drawn out of phase for clarity. 
The energy-normalized continuum wave functions $\phi^{(N^+)}_{X^+/A^+}(E,R)$ are calculated as described in Section \ref{ssec:res} for the different $N^+$ partial waves.
The corresponding Franck-Condon factors are displayed as a function of the H$^+$ + H(1s) fragment kinetic energy $E$ in the lower panels of Fig. \ref{fig4}.

\begin{figure}[bt]
\centering
\includegraphics[width=1\columnwidth]{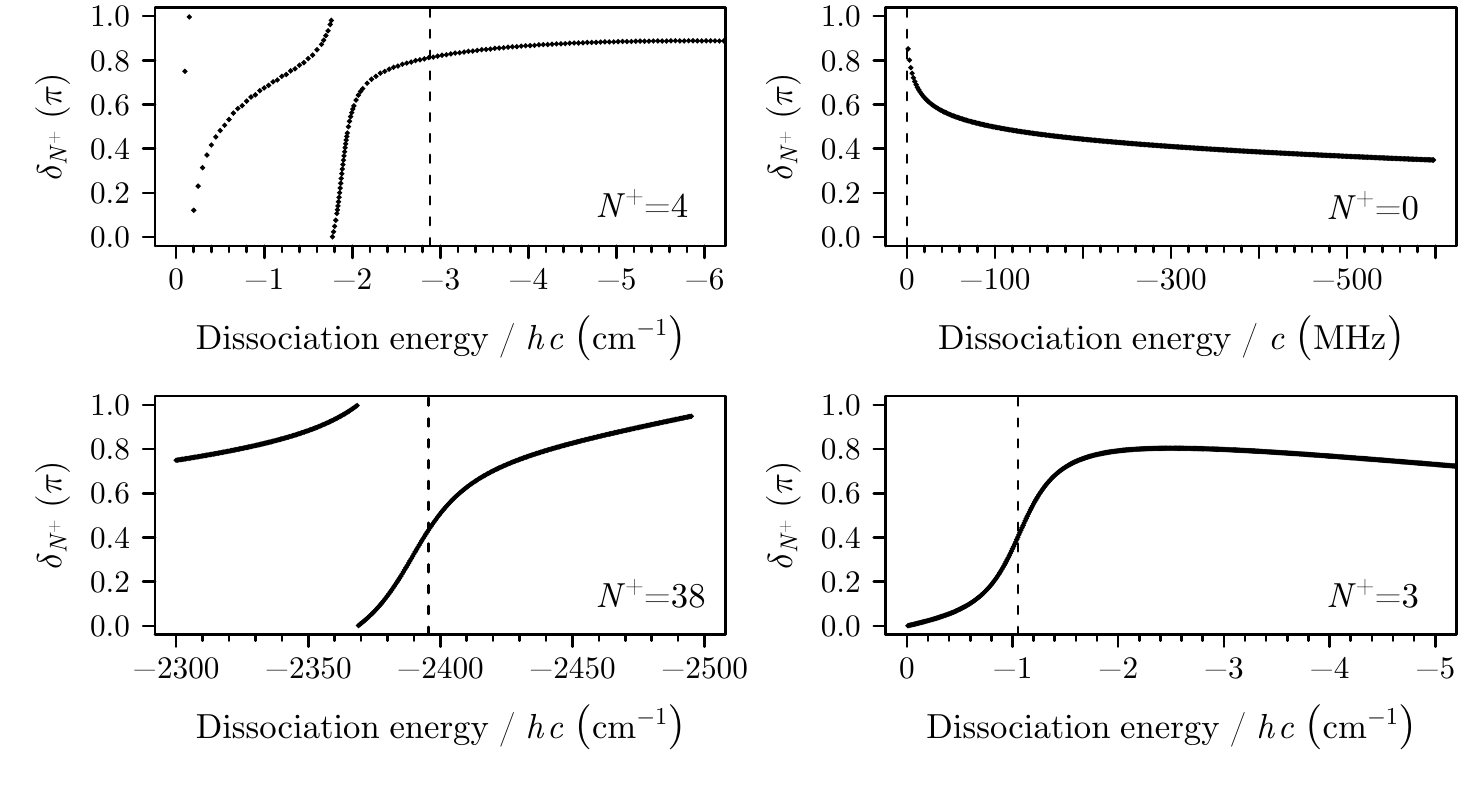}
\caption{Shape and orbiting resonances of the X$^+$ (left) and A$^+$ (right) states of H$_2^+$. Scattering phase shift $\delta_{N^+}$ modulo $\pi$ as a function of the fragment kinetic energy. The energy of the top of the centrifugal barriers are marked with vertical dashed lines.
\label{fig5}}
\end{figure}

Despite the simplification introduced by disregarding the $R$ dependence of the electronic transition moment, Fig. \ref{fig4} illustrates the following five important aspects of the dissociative-ionization behavior:

(i) The X$^+$, $N^+=4$ and $7$ cross sections are dominated by sharp and very intense features associated with the X$^+$ (18,4) and (17,7) shape resonances, respectively (see dashed lines in the corresponding panels). The scattering phase $\delta_{N^+}(E)$ reveals a jump of $\pi$ at the resonance positions, in accordance with Eq. \eqref{eq:BW} (see upper left panel in Fig. \ref{fig5} for the (18,4) resonance).

(ii) The A$^+$, $N^+=3$ cross section also reveals a resonance near threshold. However, this feature is located close to, and a substantial part of the change of the phase shift occurs above, the top of the centrifugal barrier (see lower right panel in Fig. \ref{fig5}). The overall phase shift changes by less than $\pi$.
Davis and Thorson \cite{davis78a} have called such features orbiting resonance, thus distinguishing between orbiting (resonance above barrier maximum) and shape (resonance below barrier maximum) resonances, which others often consider to be equivalent.
The behavior of the scattering phase shift serves as a distinguishing characteristic feature, being almost $\pi$ for a shape resonance and less than $\pi$ for an orbiting resonance. The lower left panel in Fig. \ref{fig5} shows the scattering phase shift near the X$^+$ (2,38) resonance, which is also located close to the maximum of the centrifugal barrier. In this case, part of the change of the phase shift occurs above the top of the barrier, but the overall phase shift equals to $\pi$, indicating for a shape resonance.

(iii) All cross sections presented in Fig. \ref{fig4} exhibit broad oscillations for both the X$^+$ and the A$^+$ contributions and return to zero between neighbouring maxima. These nodal points are the nuclear equivalent of what is known as Cooper minima in photoionization \cite{fano68a}. The oscillations are also observable in the experimental spectra (see Fig. \ref{fig2}), but the contrasts in these spectra are reduced because each spectrum represents a superposition of the contributions of several $N^+$ channels.

(iv) The low-energy onset of the dissociative-ionization cross sections gradually shifts to higher energies with increasing value of $N^+$, as expected from the Wigner threshold law \cite{fano86a}
\begin{equation}
\sigma \propto E^{N^+ + 1/2}.
\end{equation}       

(v) The A$^+$, $N^+=0$ cross section represents an exception to the Wigner threshold law, because the cross section is nonzero at threshold and first decreases with increasing energy. We attribute this exceptional behavior to the existence of the A$^+$ (1,0) level, which is located extremely close to threshold with a calculated nonrelativistic dissociation energy of 7.138596~MHz (7.139253~MHz in the calculation of Carbonell et al. \cite{carbonell03a}). The scattering phase shift, displayed in the upper right panel of Fig. \ref{fig5}, was obtained by integrating up to 6000~$a_0$ using the same potential as used for the calculation of the dissociation energy of the A$^+$ (1,0) state, thus ensuring that the increase of the phase shift does not result from an artificial box state.

Interestingly, the experimental spectra of ortho H$_2$ show a sharp line at threshold (see corresponding panels in Fig. \ref{fig2}), which might arise from the A$^+$ (1,0) state. Unfortunately, it may also be caused by a transition to the X$^+$ (19,1) level which is located 0.2207~cm$^{-1}$ below threshold.

\section{Conclusion and outlook}
The results presented in this article revealed the complexity of threshold phenomena near the dissociative-ionization limit of H$_2$. In the analysis of the experimental data, obtained by high-resolution photoelectron spectroscopy, we have focussed on the phenomena resulting from the nuclear degrees of freedom. Comparison of theory and experiment revealed a broad range of spectral features, including shape resonances, orbiting resonances and features resulting from extremely weakly bound levels, and broad energy- and $N^+$-dependent oscillations of the continuum cross sections.

In future work we intend to also consider the effect of the photoelectron (or Rydberg electron) at the three-body dissociation threshold H$^+$ + H(1s) + e$^-$. This will necessitate a complete description of the coupled nuclear and electronic motion in molecular hydrogen and the simultaneous treatment of ionization and dissociation continua \cite{jungen97a}. We indeed suspect that the reason for the discrepancy between the experimental and calculated width of the X$^+$ (17,7) shape resonance of H$_2^+$ might arise from nonadiabatic effects involving the Rydberg electron.

\section*{Acknowledgement}
We thank Dr. Ch. Jungen (Orsay) for his interest in the present work and for useful discussions and N. Hölsch for help with the fitting routines. 
This work is supported financially by the Swiss National Science Foundation under Project No. 200020-159848.

\end{document}